\documentclass[prl,amsmath,twocolumn,showpacs,superscriptaddress,letter]{revtex4}

\usepackage{amssymb}
\usepackage{graphicx}
\usepackage{natbib}

\begin{document}

%alternative title
%The Ioffe-Regel analogue of electron spin relaxation time saturation: the origin of the anomalous spin lifetime of conduction electrons in MgB$_2$
\title{Generalized Elliott-Yafet theory of electron spin relaxation in metals: the origin of the anomalous electron spin life-time in MgB$_2$}

\author{F. Simon}
\email{simon@esr.phy.bme.hu}
\affiliation{Budapest University of Technology and Economics, Institute
of Physics and Condensed Matter Research Group of the Hungarian Academy of
Sciences, H-1521 Budapest, Hungary}
\author{B. D\'{o}ra}
\affiliation{Budapest University of Technology and Economics, Institute
of Physics and Condensed Matter Research Group of the Hungarian Academy of
Sciences, H-1521 Budapest, Hungary}
\affiliation{Max-Planck-Institut f\"ur Physik
Komplexer Systeme, N\"othnitzer Str. 38, 01187 Dresden,
Germany}
\author{F. Mur\'{a}nyi$^{\dag }$}
\affiliation{Budapest University of Technology and Economics, Institute
of Physics and Condensed Matter Research Group of the Hungarian Academy of
Sciences, H-1521 Budapest, Hungary}
\author{A. J\'{a}nossy}
\affiliation{Budapest University of Technology and Economics, Institute
of Physics and Condensed Matter Research Group of the Hungarian Academy of
Sciences, H-1521 Budapest, Hungary}
\author{S. Garaj}
\affiliation{Institute of Physics of Complex Matter, FBS Swiss
Federal Institute of Technology (EPFL), CH-1015 Lausanne,
Switzerland}
\author{L. Forr\'o}
\affiliation{Institute of Physics of Complex Matter, FBS Swiss
Federal Institute of Technology (EPFL), CH-1015 Lausanne,
Switzerland}
\author{S. Bud'ko}
\affiliation{Ames Laboratory, U.S. Department of Energy and Department of
Physics and Astronomy, Iowa State University,
Ames, Iowa 50011, USA}
\author{C. Petrovic $^{\ddag}$}
\affiliation{Ames Laboratory, U.S. Department of Energy and Department of
Physics and Astronomy, Iowa State University,
Ames, Iowa 50011, USA}
\author{P. C. Canfield}
\affiliation{Ames Laboratory, U.S. Department of Energy and Department of
Physics and Astronomy, Iowa State University,
Ames, Iowa 50011, USA}

\begin{abstract}
The temperature dependence of the electron spin relaxation time in MgB$_2$ is anomalous as it does not
follow the temperature dependence of the resistivity above 150 K, it has a maximum around 400 K, and it decreases for higher temperatures. This violates the well established
Elliot-Yafet theory of electron spin relaxation in metals. We show that the anomaly occurs when the quasi-particle scattering rate (in energy units) becomes comparable to the energy difference between the conduction- and a neighboring band. We find that the anomalous behavior
 is related to the unique band structure of MgB$_2$ and the large electron-phonon coupling. The saturating spin-lattice relaxation can be
 regarded as the spin transport analogue of the Ioffe-Regel criterion of electron transport.
\end{abstract}

%74.70.Ad   Metals; alloys and binary compounds (including A15, Laves
%phases, etc.)
%74.25.Nf   Response to electromagnetic fields (nuclear magnetic
%resonance, surface impedance, etc.)
%76.30.Pk   Conduction electrons
%74.25.Ha   Magnetic properties

\maketitle

%INTRO
%\section{Introduction}

Knowledge of the electron spin-lattice relaxation time, $T_1$, of conduction
electrons plays a central role in assessing the
applicability of metals for information processing using electron
spins, spintronics \cite{FabianRMP}. $T_1$ is the time it takes
for the conduction electron spin ensemble to relax to its thermal
equilibrium magnetization after a non-equilibrium magnetization has
been induced e.g. by conduction electron-spin resonance (CESR) excitation
\cite{Feher} or by a spin-polarized current \cite{FabianRMP}.
The Elliott-Yafet (EY) theory of $T_1$ in metals \cite{Elliott,YafetReview} has been well established in the past 50 years on various systems
such as elemental metals \cite{BeuneuMonodPRB1978}, strongly correlated one-dimensional \cite{ForroSSC1982}, and some of the alkali
fulleride salt \cite{PetitPRB1996} metals.
It is based on the fact that the spin part of the conduction electron wave functions is not
a pure Zeeman state but is an admixture of the
spin up and down states due to spin-orbit (SO) coupling. As a result, momentum scattering due to phonons or impurities induces electron spin-flip, which leads to spin relaxation. Typically every millionth momentum scattering is accompanied by the electron spin-flip due to the relative weakness of the SO coupling. Thus, $T_1\gg \tau$ ($\tau$ being the momentum relaxation time) which explains the motivation behind the efforts devoted to the spintronics applications of metals.

A consequence of the EY theory is the so-called Elliott-relation, i.e. a proportionality between $T_1$ and $\tau$ \cite{Elliott}:

\begin{eqnarray}
\frac{1}{T_1}=\alpha\left(\frac{L}{\Delta E}\right)^2\frac{1}{\tau}
\label{ElliottRelation}
\end{eqnarray}

\noindent Here $\alpha$ is a band structure dependent constant and for most elemental metals $\alpha \approx 1..10$ (Ref. \cite{BeuneuMonodPRB1978}). $L$ is the SO splitting for spin up and down electrons in a valence (or unoccupied) band near the conduction band with an energy separation of $\Delta E$. E.g. in sodium, the conduction band is $3s$ derived and the relevant SO state is the $2p$ with $\Delta E=30.6\text{ eV}$ and $L=0.16\text{ eV}$ giving $(L/ \Delta E)^2=2.7\cdot 10^{-5}$ \cite{YafetReview}.

The Elliott-relation shows that the temperature dependent resistivity and CESR line-width are proportional, the two being proportional to the inverse of $\tau$ and $T_1$, respectively. This enabled to test experimentally its validity for the above mentioned range of metals. Much as the Elliott-relation has been confirmed, it is violated in MgB$_2$ as therein the CESR line-width and the resistivity are not proportional above 150 K \cite{SimonPRL2001}.

Here, we study this anomaly using MgB$_2$ samples with different B isotopes and impurity concentrations and we show that the anomalous effect is indeed intrinsic to MgB$_2$. We explain the anomaly with an exact treatment of the SO scattering of conduction electrons in the presence of a nearby band with energy separation $\Delta E$, by extending the Mori-Kawasaki formula developed for localized spins to itinerant electrons. The result shows that the Elliott-relation breaks down when $\Delta E$ is comparable to $\hbar/\tau$. Adrian deduced a similar result with a qualitative argument \cite{AdrianPRB1996}.

The role of $\Delta E$ is disregarded in the EY theory since typical values are $\Delta E\approx 10\text{ eV}$ and $\hbar/\tau=2 \pi k_{\text{B}} T \lambda \approx 6\text{ meV}$ at $T=100 \text{ K}$ and $\lambda=0.1$ electron-phonon coupling. We show that the occurrence of the anomaly in MgB$_2$ is related to the unique features in its band structure and the large electron-phonon coupling.

%\section{Experimental}

%EXPERIMENTAL
We performed CESR measurements on three kinds of
fine powder MgB$_2$ with isotope pure $^{10}$B, $^{11}$B, and natural boron (20 \% $^{10}$B and 80 \% $^{11}$B). The samples have slightly different impurity content, shown by the varying residual CESR line-width, $\Delta B_0$. The temperature dependent $T_1$ and the CESR line-width, $\Delta B$, are related: $\Delta B=\Delta B_0+1/\gamma T_1$, where $\gamma/2 \pi=28$ GHz/T is the electron gyromagnetic factor. ESR spectroscopy was done on a Bruker X-band spectrometer (center field 0.33 T) in the 4-700 K temperature range on samples sealed under He in quartz tubes. The most important result of the current report, the anomalous temperature dependence of $\Delta B$ or $T_1$, is independent of sample morphology, isotope content, or thermal history. $\Delta B$ is also independent of the magnetic field, apart from a small change in $\Delta B_0$ \cite{SimonMgB2PRB2007}. Resistance on pellet samples and SQUID magnetometry were studied on the same batch as those used for ESR. The $RRR>20$ and the sharp ($< 0.5$ K) superconducting transition attest the high quality of the samples. Heating the samples in the ESR measurement (about 1 h duration) to 700 K does not affect the superconducting properties as shown by magnetization measurements.

%Results and discussion
%\section{Results}

\begin{figure}
\includegraphics[width=0.9\hsize]{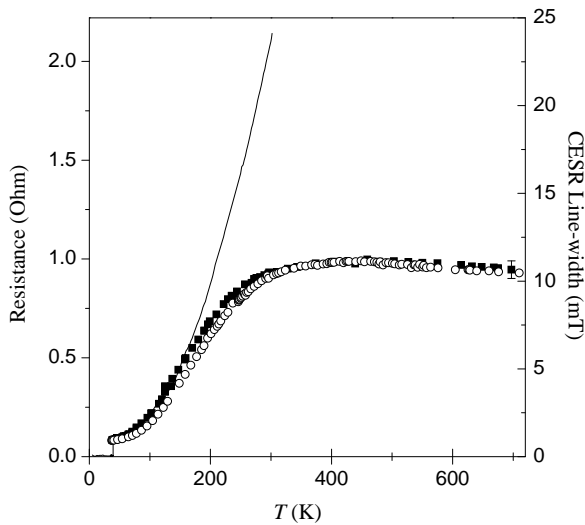}
\caption{Comparison of the temperature dependent CESR line-width ($\blacksquare$: Mg$^{11}$B$_2$, $\bigcirc$:MgB$_2$ of natural boron) and the resistance (solid curve) for Mg$^{11}$B$_2$. The two types of data overlap in the 40-150 K temperature range. A representative error bar is shown.}
\label{Fig1_ESR_linewidth_resistance}
\end{figure}

We reported previously the anomalous temperature dependence of
the CESR line-width in Mg$^{11}$B$_2$: although the line-width follows the resistance for the 40-150 K temperature range, it deviates above 150 K and saturates above 400 K \cite{SimonPRL2001}. This was confirmed independently \cite{MonodPrivComm,RettoriPRL2002}. To our knowledge, this is the only known metal where such phenomenon is observed. We extended the previous measurement to 700 K and the result is shown in Fig. \ref{Fig1_ESR_linewidth_resistance}. Interestingly, the CESR line-width does not just saturate at high temperatures, as found previously, but \emph{decreases} slightly above 500 K. The result is reversible upon cooling with no dependence on the thermal treatment protocol. The phenomenon is reproduced on several samples of different purity and boron isotopes, thus it is intrinsic to MgB$_2$.

We explain the anomalous temperature dependence of $T_1$ in general before including the specifics of MgB$_2$. The Elliott-Yafet theory disregards the magnitude of $\tau$ and takes life-time effects only to lowest order into account \cite{Elliott,YafetReview}. The extended description involves the Kubo-formalism and is based on a two-band model Hamiltonian, $H=H_0+H_{\text{SO}}$, where:

%\begin{widetext}

%old version
%\begin{gather}
%H_0=\sum_{k,\nu,s}\left[\epsilon_\nu(k)+\hbar \gamma B
%s\right]c^+_{k,\nu,s}c_{k,\nu,s}+H_{\text{scatt}},\\
%H_{\text{SO}}=\sum_{k,\nu\neq\nu',s,s'} L_{s,s'}(k)c^+_{k,\nu,s}c_{k,\nu',s'}
%\label{ModelHamiltonian}
%\end{gather}

\begin{equation}
\begin{split}
H_0=\sum_{k,\nu,s}\left[\epsilon_\nu(k)+\hbar \gamma B
s\right]c^+_{k,\nu,s}c_{k,\nu,s}+H_{\text{scatt}},\\
H_{\text{SO}}=\sum_{k,\nu\neq\nu',s,s'} L_{s,s'}(k)c^+_{k,\nu,s}c_{k,\nu',s'}
\label{ModelHamiltonian}
\end{split}
\end{equation}

%\end{widetext}

\noindent Here $\nu, \nu'=1 \text{ or }2$ are the band, $s,s'$ are spin indices, $L_{s,s'}$ is the SO coupling, and $B$ is the
magnetic
field along the $z$ direction. $H_{\text{scatt}}$ is
responsible for the finite $\tau$.
The SO coupling does not split spin up and down states in the same band for a crystal with
inversion symmetry, however it joins different spin states in the two bands \cite{FabianRMP}. The Hamiltonian in
Eq. \ref{ModelHamiltonian} is essentially the same as that considered by Elliott \cite{Elliott}. However, instead of a time-dependent perturbation treatment, we calculate $T_1$ from the Mori-Kawasaki formula \cite{mori,oshikawa}:

\begin{gather}
\frac{1}{T_1}=-\frac{1}{2\chi_0 B}\textmd{Im}G^R_{PP^+}(\omega_{\text{L}}),
\end{gather}

\noindent where $\chi_0$ is the static magnetic susceptibility, $\omega_{\text{L}}=\gamma B$ is the Larmor frequency, and $G^R_{PP^+}(\omega)$ is the Fourier transform of

%\begin{gather}
%G^R_{PP^+}(t)=-i\Theta(t)\langle[P(t),P^+(0)]\rangle_{H_0}\\
%P=[H_{SO},S^+],
%\label{FourierTransform}
%\end{gather}

\begin{equation}
\begin{split}
G^R_{PP^+}(t)&=-i\Theta(t)\langle[P(t),P^+(0)]\rangle_{H_0},\\
P&=[H_{\text{SO}},S^+].
\label{FourierTransform}
\end{split}
\end{equation}

\noindent The expectation value in Eq. \ref{FourierTransform}. is evaluated with the unperturbed Hamiltonian, $H_0$.

Assuming that the two bands are separated by $\Delta E(k)=\epsilon_1(k)-\epsilon_2(k)=\hbar \Delta\omega(k)$, a standard calculation
yields \cite{klasszikus}:

\begin{gather}
\frac{1}{T_1}= \left\langle\frac{ L_z^2(k_F)+2|L_{\downarrow,\uparrow}(k_F)|^2}{\hbar^2}
\frac{\tau}{1+(\Delta \omega(k_F)\tau)^2}\right\rangle,
\label{ESRT1}
\end{gather}

\noindent where the $\langle\dots\rangle$ means Fermi surface averaging,
$L_z(k)=L_{\uparrow,\uparrow}(k)-L_{\downarrow,\downarrow}(k)$,
and we neglected $\omega_{\text{L}}$, which
is small compared to $\Delta \omega(k_F)$.
Eq. \ref{ESRT1}. was previously deduced by Adrian using a qualitative argument, which involved an effective magnetic field, $L/\hbar \gamma$, fluctuating with $\tau$ correlation time due to the SO coupling \cite{AdrianPRB1996}.

We approximate Eq. \ref{ESRT1} using effective values for the band-band energy separation and the
SO coupling:

\begin{gather}
\frac{1}{T_1}= \frac{L_{\text{eff}}^2}{\hbar^2}
\frac{\tau}{1+\Delta \omega_{\text{eff}}^2\tau^2},
\label{ESRT1_approximated}
\end{gather}

\noindent This result returns the Elliott-relation when $ \tau \Delta \omega_{\text{eff}}\gg 1$ and gives a decreasing spin relaxation rate with increasing $\tau^{-1}$ when $\tau \Delta \omega_{\text{eff}} \leq 1$, thus it can be regarded as a generalization of the Elliott-Yafet theory. In the following, we show that it describes the spin relaxation in MgB$_2$.

\begin{figure}
\includegraphics[width=1.0\hsize]{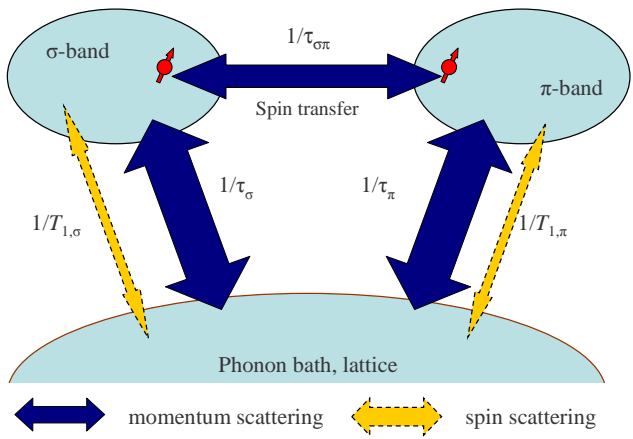}
\caption{(Color online) Schematics of the spin-lattice relaxation in MgB$_2$ in the two-band model framework. The arrow thicknesses represent the relaxation rates (not to scale). Note that the inter-band momentum scattering rate is larger than the spin-lattice relaxation rates, therefore there is a spin transfer between the two types of bands.
}
\label{Fig2_two_band_model_schematics}
\end{figure}

Electronic properties of MgB$_2$ are described by the so-called two-band model meaning that the conduction bands related to the boron $\sigma$ and $\pi$ bonds have different electron-phonon couplings, different affinity to defects, and that the inter-band momentum scattering is weaker than the intra-band ones \cite{MazinPRL2002}. As a result, the conductivity is given by a parallel
resistor formula \cite{MazinPRL2002}, i.e. the band with
longer $\tau$ dominates the transport. In
contrast, the CESR spin relaxation is dominated by the
band with \emph{shorter} $T_1$. Although
the inter-band momentum scattering time, $\tau_{\sigma \pi}$ is longer than the
intra-band momentum scattering times, $\tau_{\sigma}$ and $\tau_{\pi}$, it is still much shorter than
$T_1$. Thus an
electron with a given spin state is scattered back and forth between
the two types of bands several times before flipping its spin, which is depicted in Fig. \ref{Fig2_two_band_model_schematics}. The overall $1/T_1$ is the average of the spin-lattice
relaxation rates weighted by the relative DOS on the $\sigma$ and $\pi$ bands,
$N_{\pi}=0.56$ and $N_{\sigma}=0.44$ \cite{LouieNAT2002}:

\begin{equation}
\frac{1}{T_1}=
\frac{N_{\pi}}{T_{1,\pi}}+\frac{N_{\sigma}}{T_{1,\sigma}}
\label{ComposedRelaxation}
\end{equation}

\begin{figure}
\includegraphics[width=1.0\hsize]{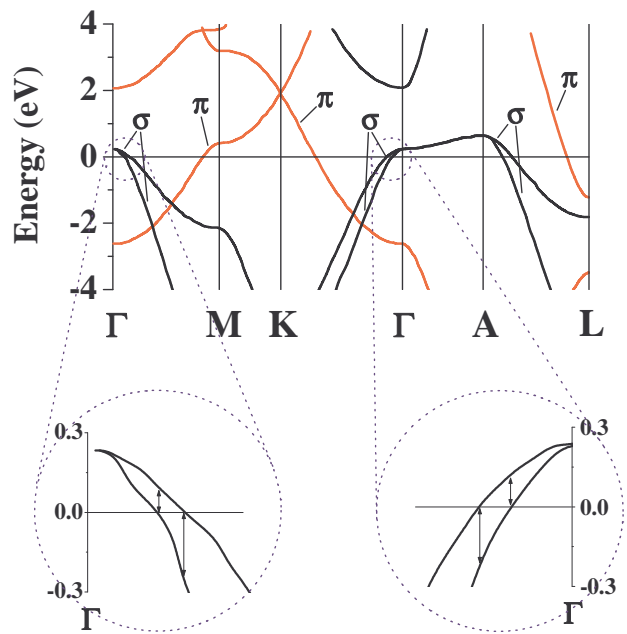}
\caption{(Color online). Band structure of MgB$_2$ near the Fermi energy after Refs. \cite{MazinPRL2001,MedvedevaPRB2001}.
Two of the $\sigma$ bands (black) cross the Fermi surface near each other in the vicinity of the $\Gamma$ and A points, whereas $\pi$
bands (red) are separated from other bands with a larger optical gap at the crossing. We also show the dispersion with 8 times larger wave-vector resolution around the $\Gamma$ points with vertical arrows for possible $\Delta E_{\sigma}$ values.}
\label{Fig3_DOS}
\end{figure}

In Fig. \ref{Fig3_DOS}., we show the band structure of MgB$_2$ from Refs. \cite{MazinPRL2001,MedvedevaPRB2001} near the Fermi energy. Two boron $\sigma$ and two $\pi$ bands cross the Fermi energy such that the $\pi$ bands are well separated from other bands with $\Delta E_{\pi}\geq 2 \text{ eV}$ whereas the two $\sigma$ bands are close to each other and $\Delta E_{\sigma} \approx 0.2 \text{ eV}$. Based on the above theory and Eq. \ref{ESRT1_approximated}., we conclude that $T_1$ follows the EY mechanism for the $\pi$ bands, whereas it is described the by the novel mechanism for the $\sigma$ bands. With this in mind and the two band model result of Eq. \ref{ComposedRelaxation}, we describe the CESR line-width with:

\begin{equation}
\Delta B=\Delta B_0+ \frac{1}{\gamma\hbar^2}\left(\frac{N_{\pi}L_{\text{eff,}\pi}^2}{\Delta \omega^2_{\text{eff,}\pi}}\frac{1}{\tau_{\pi}}+\frac{N_{\sigma}L_{\text{eff,}\sigma}^2\tau_{\sigma}}{1+\Delta \omega_{\text{eff,}\sigma}^2 \tau_{\sigma}^2}
\right)
\label{TwoBandAdrianModelSimplified}
\end{equation}

\noindent where we introduced the band index for the parameters. The momentum relaxation times are calculated using the Debye-model and assuming clean samples, i.e. zero residual scattering:

\begin{eqnarray}
\frac{1}{\tau_{n}}=\frac{2\pi
\text{k}_{\text{B}}T \lambda_{\text{tr}, n}}{\hbar } \mathop{
\int_0^{\omega_{\text{D}}}}\frac{d\Omega}{\Omega}\left(\frac{\Omega}{\omega_{\text{D}}}
\right)^4\left[\frac{\hbar\Omega/\text{k}_{\text{B}}T}{\sinh
\frac{\hbar \Omega}{2 \text{k}_{\text{B}}T} } \right]^2,
\label{Tau2BandDebye}
\end{eqnarray}

\noindent where $n=\sigma, \pi$, $\omega_{\text{D}}$ is the Debye frequency, and
$\lambda_{\text{tr}, n}$ are the transport electron-phonon
couplings from Ref. \cite{MazinPRL2002}, which contain both intra- and inter-band scattering.

\begin{figure}
\includegraphics[width=1.0\hsize]{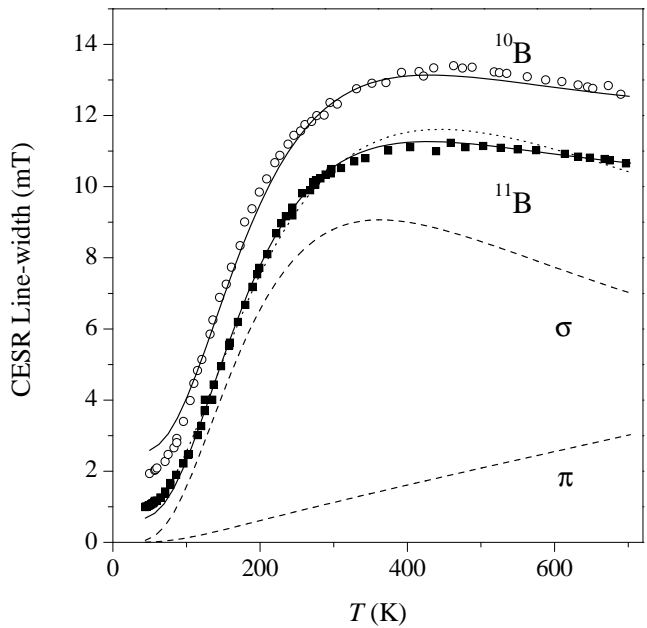}
\caption{Measured (symbols) and calculated (solid curves) CESR line-width in MgB$_2$ with $^{11}$B and $^{10}$B. Note the larger residual line-width in the latter sample.
Dashed curves show the contributions to the line-width from the $\sigma$ and $\pi$ bands for the $^{11}$B. Dotted curve shows a calculation for the the $^{11}$B sample assuming $1/T_1$ is due to $\sigma$ bands only.}
\label{Fig4_fit}
\end{figure}

In Fig. \ref{Fig4_fit}., we show the CESR line-width for Mg$^{11}$B$_2$ and Mg$^{10}$B$_2$ between 40 and 700 K and the calculated line-width using Eq. \ref{TwoBandAdrianModelSimplified}. with parameters in Table \ref{Fit_Parameters}. obtained from a fit. Results on the natural boron sample are identical to the data on the Mg$^{11}$B$_2$ within experimental error and are not shown. The larger residual line-width in the $^{10}$B ($\Delta B_0=2\text{ mT}$) than in the $^{11}$B sample ($\Delta B_0=1\text{ mT}$) is related to a larger defect concentration in the starting boron, the preparation method and the starting Mg being identical. Apart from this, the only difference between the two samples are the different Debye temperature, $\Theta_{\text{D}}$. The calculated CESR line-width (solid curves) reproduces well the experimental data with the parameters in Table \ref{Fit_Parameters}. The dotted curve in Fig. \ref{Fig4_fit}. is a calculation assuming that relaxation is given by the $\sigma$ bands alone, which accounts relatively well for the data with three free parameters ($L_{\sigma}$, $\Delta E_{\text{eff},\sigma}$, and $\Delta B_0$). However, it fails to reproduce the slope of $\Delta B$ at higher temperatures, which shows the need to include relaxation due to the $\pi$ bands.

The determination of $\Delta E_{\text{eff},\sigma}\approx 0.2 \text{ eV}$ is robust as it is given by the temperature where the maximal $\Delta B$ is attained and its value is close to values expected from the band structure (arrows in Fig. \ref{Fig3_DOS}.). Knowledge of $\Delta E_{\text{eff},\sigma}$ allows to determine the SO splitting independently, $L_{\text{eff},\sigma}=0.64 \text{ meV}$, as usually only the $L/\Delta E$ ratio is known. The SO splitting for the atomic boron $2p$ orbital is $L=0.23\text{ meV}$ (Ref. \cite{YafetReview}), which is in a reasonable agreement with the experimental value. $\Delta E_{\pi}$ was fixed to 2 eV which affects $L_{\text{eff},\pi}$ as these are not independent.

The isotope effect on $\Theta_{\text{D}}$ is $^{10}\Theta_{\text{D}}/^{11}\Theta_{\text{D}}=1.04$, that is close to the expected $\sqrt{11/10}$ ratio. The $\Theta_{\text{D}}$ values are in agreement with the 440..1050 K values in the literature, which scatter depending on the experimental method \cite{GasparovJETP2001,PuttiEPJB2002}. We note that the model could be improved by including the Einstein model of phonons or by an exact treatment of the band structure dependent SO coupling \cite{FabianPRL1999}, and band-band separation.

\begin{table}[tbp]
\caption{Parameters used to calculate the CESR line-width in MgB$_2$. The given standard deviations indicate the free parameters of the fit.}
%\begin{tabular}{llllllllllll}
\begin{tabular*}{0.5\textwidth}{@{\extracolsep{\fill}}cccccccccccc}
\hline \hline
 \multicolumn{2}{c}{$\lambda_{\text{tr}}$}\cite{MazinPRL2002} &  \multicolumn{2}{c}{$L_{\text{eff}}$ (meV)}&  \multicolumn{2}{c}{$\Delta E_{\text{eff}}$ (eV)} &
 \multicolumn{2}{c}{$\Theta_{\text{D}}$ (K)} \\
$\sigma$& $\pi$&$\sigma$ & $\pi$ &  $\sigma$ & $\pi$ & $^{11}B$ & $^{10}B$ \\
\hline
 1.09 & 0.46 &  0.64(2) & 2.8(1) & 0.194(5)&2 & 535(15) & 555(15)\\
\hline \hline
\label{Fit_Parameters}
\end{tabular*}
\end{table}

Finally, we note that the maximum of $1/T_1$ occurs when $\tau\Delta \omega \approx 1$. This coincides with the Ioffe-Regel criterion for the electron transport \cite{GunnarssonReviewTransport} when the band-band separation is comparable to the bandwidth, $w$, e.g. in narrow band metals. For MgB$_2$, $w \approx 10\text{ eV}$ \cite{MazinPRL2001} therefore saturation of the CESR line-width is not accompanied by a saturation of electrical resistivity.

In conclusion, we explained the anomalous spin-lattice relaxation in MgB$_2$ by extending the Elliott-Yafet theory to the case of rapid momentum scattering and near lying bands. The anomaly does not occur in conventional metals, which have small electron-phonon coupling and well separated bands. A similar phenomenon, the so-called Dyakonov-Perel relaxation \cite{FabianRMP}, occurs for semiconductors without inversion symmetry, although its physical origin is different.
The band structure of some of the other diborides in e.g. BeB$_2$ and CaB$_2$ predicts \cite{MedvedevaPRB2001} similar phenomena but conventional spin relaxation in AlB$_2$, ScB$_2$, and YB$_2$. We also predict that the described effect is sensitive to pressure since this shifts the $\sigma$ bands \cite{KobayashiJPSJ2001}.

%CONCLUSION

%\section{Acknowledgements}
We are grateful to J. Fabian and A. Virosztek for enlightening discussions.
FS and FM acknowledge the Bolyai programme of the Hungarian Academy of
Sciences and the Humboldt Foundation for support. Work
supported by the Hungarian State Grants (OTKA) No. F61733,  K72613, and NK60984. Ames Laboratory is operated for the U.S. Department of
Energy by Iowa State University under Contract No. W-7405-Eng-82.

$^{\ast }$ Corresponding author: simon@esr.phy.bme.hu

$^{\dag }$ Present address: IFW Dresden, Institute
for Solid State Research, P. O. Box 270116, D-01171 Dresden, Germany

$^{\ddag}$ Present address: Condensed Matter Physics and Materials Science Department, Brookhaven National
Laboratory, Upton, New York 11973-5000, USA

%\bibstyle{apsrev}
%\bibliography{MgB2}

\end{document}